\documentclass[12pt]{article}

\pdfoutput=1

\usepackage[english]{babel}
\usepackage{fancyhdr}
\usepackage{amsmath}
\usepackage{amssymb}
\usepackage{amsfonts}
\usepackage{psfrag}
\usepackage[applemac]{inputenc}
\usepackage{graphicx,wrapfig}
\usepackage{pstricks}
\usepackage{mathrsfs}
\usepackage{slashed}            
\usepackage{subfig}             
\usepackage{xspace}				
\usepackage[sort&compress,square,comma,numbers]{natbib}
\usepackage{float}
\usepackage{dcolumn}
\usepackage{bm}
\usepackage{epstopdf}
\usepackage{feynmp}
\usepackage{array}
\usepackage{multirow}
\newcolumntype{?}{!{\vrule width 1pt}}
\newcolumntype{Y}{>{\raggedright\arraybackslash}X} 

\usepackage{booktabs,tabularx}

\DeclareFontFamily{OT1}{pzc}{}
\DeclareFontShape{OT1}{pzc}{m}{it}{<-> s * [1.100] pzcmi7t}{}
\DeclareMathAlphabet{\mathpzc}{OT1}{pzc}{m}{it}

\newcommand{\beq}{\begin{eqnarray}}
\newcommand{\eeq}{\end{eqnarray}}
\newcommand{\bea}{\begin{eqnarray}}
\newcommand{\eea}{\end{eqnarray}}
\newcommand{\bag}{\begin{align}}
\newcommand{\eag}{\end{align}}

\makeatletter
\newcommand{\thickhline}{%
    \noalign {\ifnum 0=`}\fi \hrule height 1pt
    \futurelet \reserved@a \@xhline
}
\newcolumntype{"}{@{\hskip\tabcolsep\vrule width 1pt\hskip\tabcolsep}}
\makeatother

\usepackage[
	colorlinks=true,
	citecolor=black,
	linkcolor=black,
	urlcolor=blue,
	hypertexnames=false]{hyperref}
	
\newcommand{\arXiv}[2]{\href{http://arxiv.org/pdf/#1}{{\tt #2/#1}}}
\newcommand{\arXivold}[1]{\href{http://arxiv.org/pdf/#1}{{\tt #1}}}

\begin{document}

\baselineskip=18pt

\setcounter{footnote}{0}
\setcounter{figure}{0}
\setcounter{table}{0}


\begin{titlepage}

\begin{center}
  \begin{LARGE}
    \begin{bf}
Gluon vs. Photon Production of a 750 GeV Diphoton Resonance
\vspace*{0.2cm}

   \end{bf}
  \end{LARGE}
\end{center}
\vspace{0.1cm}
\begin{center}
\begin{large}
{\bf Csaba Cs\'aki$^a$, Jay Hubisz$^b$, Salvator Lombardo$^a$, John Terning$^c$ \\}
\end{large}
  \vspace{0.5cm}
  \begin{it}

\begin{small}
$^{(a)}$Department of Physics, LEPP, Cornell University, Ithaca, NY 14853, USA
\vspace{0.2cm}\\
$^{(b)}$Department of Physics, Syracuse University, Syracuse, NY 13244, USA
\vspace{0.2cm}\\

$^{(c)}$Department of Physics, University of California, Davis, CA 95616
 \vspace{0.1cm}

\end{small}

\end{it}
\vspace{.5cm}

{\tt csaki@cornell.edu, jhubisz@syr.edu, sdl88@cornell.edu, jterning@gmail.com}

\end{center}

\vspace*{0.5cm}

\begin{abstract}
\medskip
\noindent 
The production mechanism of a 750 GeV diphoton resonance, either via gluon or photon fusion, can be probed by studying kinematic observables in the diphoton events. We perform a detector study of the two production modes of a hypothetical scalar or tensor diphoton resonance in order to characterize the features of the two scenarios. The nature of the resonance production can be determined from the jet multiplicity, the jet and diphoton rapidities, the rate of central pseudorapidity gaps, or the possible detection of forward protons from elastic photoproduction for events in the signal region. Kinematic distributions for both signals and expected irreducible diphoton background events are provided for comparison along with a study of observables useful for distinguishing the two scenarios at an integrated luminosity of 20 fb$^{-1}$. We find that decay photons from a 750 GeV scalar resonance have a preference for acceptance in the central detector barrel, while background events are more likely to give accepted photons in the detector end caps. This disfavors the interpretation of the large number of excess events found by the the Run-2 CMS diphoton search with one photon detected in the end cap as a wide spin-0 resonance signal. However, one expects more end cap photons in the case of spin-2 resonance. 
 
\end{abstract}

\bigskip

\end{titlepage}

\section*{Introduction} 

 While most interpretations of the 750 GeV diphoton excess recently reported by ATLAS and CMS \cite{ATLASdiphoton, CMS:2015dxe} assume gluon production (for a sample of such studies, see for example~\cite{Harigaya:2015ezk,Knapen:2015dap,Franceschini:2015kwy,McDermott:2015sck,Falkowski:2015swt,Agrawal:2015dbf,Bellazzini:2015nxw,Curtin:2015jcv}), 
a more minimal scenario via photoproduction (as proposed in~\cite{Csaki:2015vek, Fichet:2015vvy, Altmannshofer:2015xfo}) is also plausible. The photoproduction cross section for a 750 GeV resonance is able to account for the entire excess provided the resonance has sizable $\mathcal{O}(1\%)$ branching ratio to two photons. In this paper, we simulate the two production scenarios, provide the full cross sections taking into account both elastic and inelastic photoproduction, and present kinematical distributions useful for discriminating between the two scenarios for the production of the hypothetical resonance. Complementary to this paper, Ref.~\cite{Gao:2015igz} has studied quark-antiquark annihilation production of the resonance.

The production process of a 750 GeV diphoton resonance has observable features in the kinematical distribution of events in the signal region. In particular, the color flow of photon fusion ($\gamma \gamma$F) or $W/Z$ fusion processes would suppress the production of central hadronic activity resulting in fewer central jets compared to a gluon fusion ($gg$F) signal while also enhancing the number of central pseudorapidity gaps, central regions of the detector absent of hadronic activity. The kinematic properties of the events in the region of the diphoton excess are reported to have no significant difference compared to events above and below the excess diphoton invariant mass region. We find that this is consistent at 3.2 fb$^{-1}$ of data with a resonance produced dominantly through either $\gamma\gamma$F or $gg$F, assuming the background makes approximately half of the detected events as suggested by the number of diphoton events predicted and observed in the excess region by ATLAS~\cite{ATLASdiphoton}. However, the additional jet multiplicity expected from a $gg$F signal compared to the expected jet multiplicity from the dominant irreducible diphoton background events would already suggest an approximately 1.5-$\sigma$ excess in the total number of jets in the signal region $690 < m_{\gamma \gamma} < 810$ GeV for the 38 events measured by ATLAS if the resonance were produced though $gg$F. Furthermore, the measurement of the central rapidity gap rate, expected from the signal $\gamma \gamma$F and background events but exponentially suppressed in $gg$F events, could discriminate between a VBF or $gg$F signal but requires special care in order to ensure central tracks originating from pileup do not bury the signature. 

Due to differences in the shape of parton distribution functions and the kinematics between $\gamma \gamma$F and $gg$F events, the resonance tends to be produced more forward in $\gamma\gamma$F production. We note that this affects the rapidity distribution of the photons, preferring rapidities corresponding to the detector end caps more often than gluon fusion produced events, although this effect alone is not 
sufficiently large to fully account for the large end cap excess in the CMS diphoton search \cite{CMS:2015dxe}. We find that photons from a decaying heavy scalar resonance are more often central than photons from background events and tend to be detected in the detector barrel rather than the end caps. If the hypothetical resonance has a large width and the large excess in the 700 GeV bin of the CMS search is to be considered a signal, then this result is in tension with the interpretation of the large fraction of CMS diphoton events with one photon detected in the barrel and one detected in the end cap (EBEE category) as a scalar resonance signal. The situation for a wide resonance is improved for a spin-2 resonance for which one expects more events in the EBEE category than for a scalar signal.

Elastic photoproduction events result in forward and backward protons which can be detected by forward detectors installed by ATLAS and CMS~\cite{CERN-LHCC-2011-012,Albrow:1753795}. Elastic production is suppressed with respect to inelastic. However, the detection of two intact protons in the final state, with $m_{pp}$ matched to $m_{\gamma \gamma}$, can be used to remove background. It was estimated in~\cite{Fichet:2015vvy} that approximately 20 fb$^{-1}$ is needed for a 5-$\sigma$ discovery in this channel. In this paper we use this luminosity as a benchmark to characterize which features of the production mechanism may be apparent in the kinematic properties of excess events at or before 20 fb$^{-1}$ of data.

\section*{Production via Photon Fusion} 

\begin{figure}[t!]
\center
\includegraphics[width=.8\textwidth]{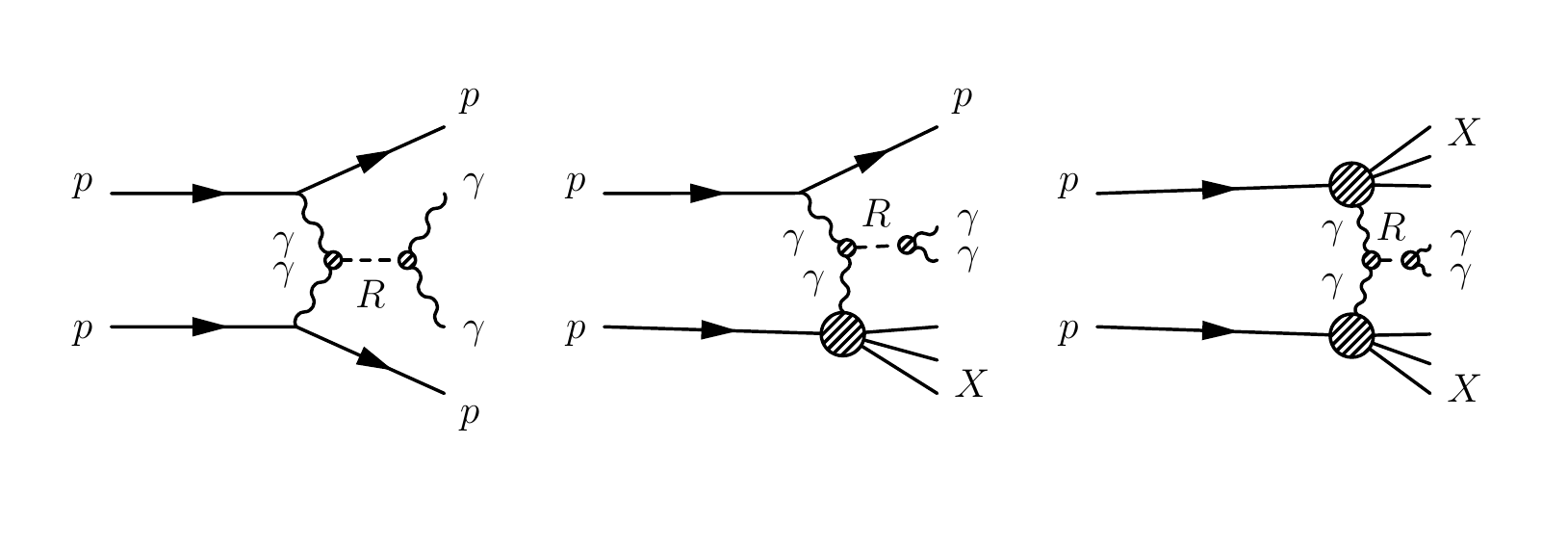}
\caption{Elastic-elastic, elastic-inelastic and inelastic-inelastic contributions to the photoproduction of the resonance $R$. } \label{fig:photoproduction}
\end{figure}

Following \cite{Csaki:2015vek,Fichet:2015vvy}, we will consider a model with an additional scalar particle $R$ with mass $m \approx 750$ GeV whose only sizable coupling to SM particles is to photons via the operator 
\begin{equation}
\frac{c_{\gamma\gamma}}{v} R F^2\ ,
\label{eq:operator}
\end{equation}
with $v=246$ GeV introduced to have dimensionless couplings, resulting in a partial width to photons $\Gamma_{\gamma\gamma}$ of 
\begin{equation}
\Gamma_{\gamma\gamma} = \frac{c_{\gamma\gamma}^2}{4\pi} \frac{m^3}{v^2}\ .
\label{eq:width}
\end{equation}
 
 \begin{figure}[t!]
\center
\includegraphics[width=.65\textwidth]{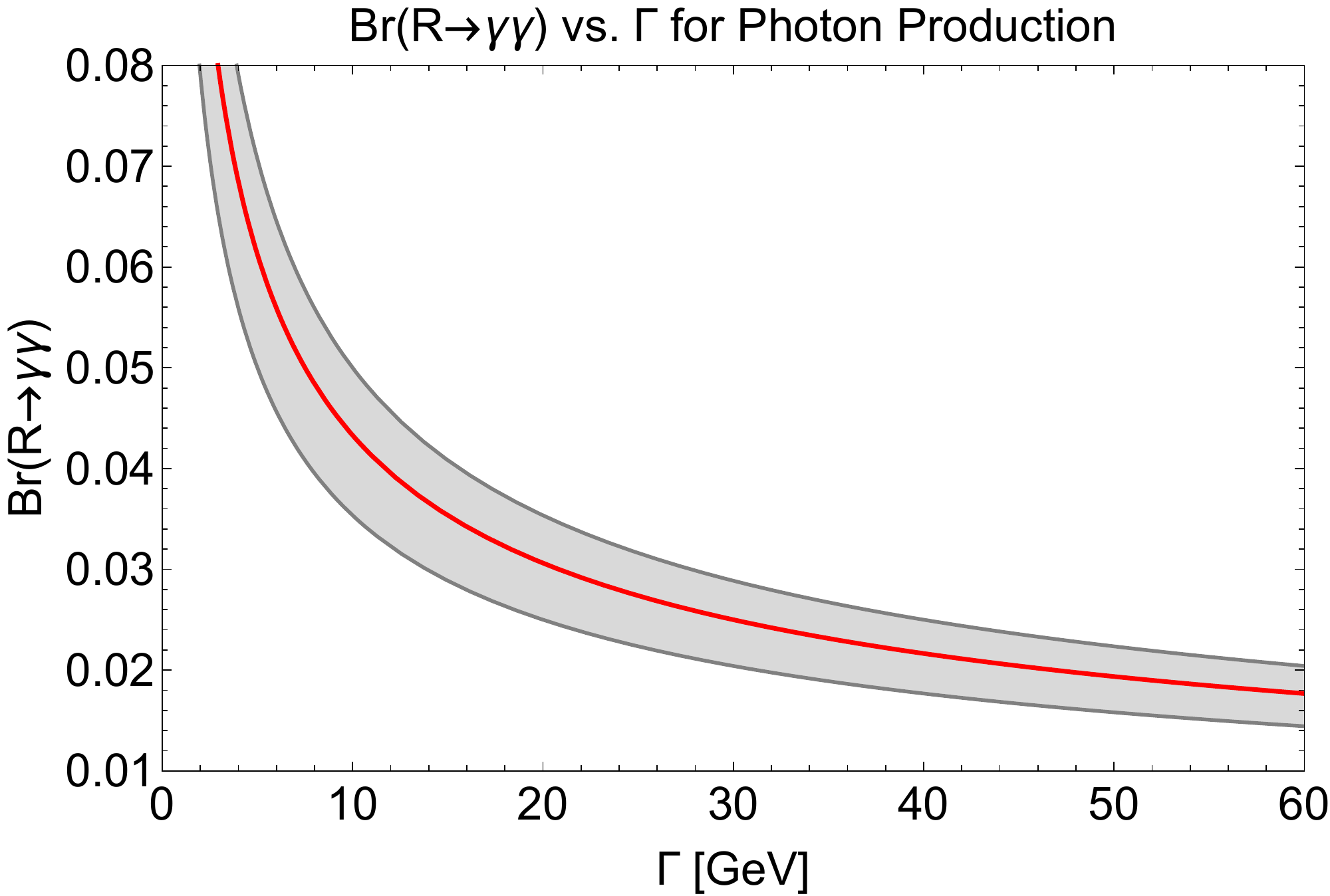}
\caption{The required relationship between the branching ratio of $R \rightarrow \gamma \gamma$ and total width $\Gamma$ to match the observed event rate, varied between 3-6 fb, assuming photon fusion dominates. The central value (red) corresponds to a 4.5 fb production cross section.} \label{fig:BRvsGamma}
\end{figure}
 
In this paper, we include the contributions from inelastic-inelastic, elastic-inelastic, and elastic-elastic processes (see Fig.~\ref{fig:photoproduction}). In the narrow width approximation, the total photoproduction cross section at $\sqrt{s} = 13$ TeV is 
\begin{equation}
\sigma_{13~\text{TeV}} = 10.8~\text{pb}~\left(\frac{\Gamma}{45~\text{GeV}} \right) \text{Br}^2(R \rightarrow \gamma\gamma),
\end{equation}
determined at leading order (LO) from {\tt MadGraph 5}~\cite{Alwall:2014hca} with the parton distribution function set {\tt NN23LO1}~\cite{Ball:2013hta}. For elastic collisions, the equivalent photon approximation is made with the improved Weizsacker-Williams formula~\cite{Frixione:1993yw} in order to account for the electromagnetic form factor of the proton. Inelastic collisions dominate the production followed by partially elastic and elastic collisions in the ratios 63:33:4, respectively. Here, we see that the rate of $\sigma(pp \rightarrow R + Z) \text{BR}(R\rightarrow \gamma \gamma) \sim 3$-$6$ fb (corresponding to the excess number of events observed by ATLAS) can be accommodated by a photoproduced resonance with total width of $45$ GeV, motivated by the best-fit width of the ATLAS excess, with branching ratio to two photons of approximately 2\%. If we allow the total width $\Gamma$ to vary, the relationship between $\text{Br}^2(R \rightarrow \gamma\gamma)$ and $\Gamma$ is fixed by matching the observed event rate of the excess and is shown in Fig.~\ref{fig:BRvsGamma}. We assume $\Gamma = 45$ GeV for the remainder of this paper, although the conclusions still apply for a narrower resonance.

The total production cross section at $\sqrt{s} = 8$ TeV is 
\begin{equation}
\sigma_{8~\text{TeV}} = 5.5~\text{pb}~\left(\frac{\Gamma}{45~\text{GeV}} \right) \text{Br}^2(R \rightarrow \gamma\gamma)\ .
\end{equation}
The ratio $\sigma_{13~\text{TeV}}/\sigma_{8~\text{TeV}}$ determined from {\tt MadGraph} is approximately 2 and does not provide an explanation for the absence of a signal in Run I diphoton searches.  However, the ratio for elastic and partially elastic production depends strongly on the finite size effects of the proton, or equivalently the maximum fraction of the proton momentum transferred to an emitted photon, and can be larger than 2 depending on the correct value~\cite{Csaki:2015vek, Fichet:2015vvy}.

\section*{Production via Gluon Fusion} 

\begin{figure}[t!]
\center
\includegraphics[width=.65\textwidth]{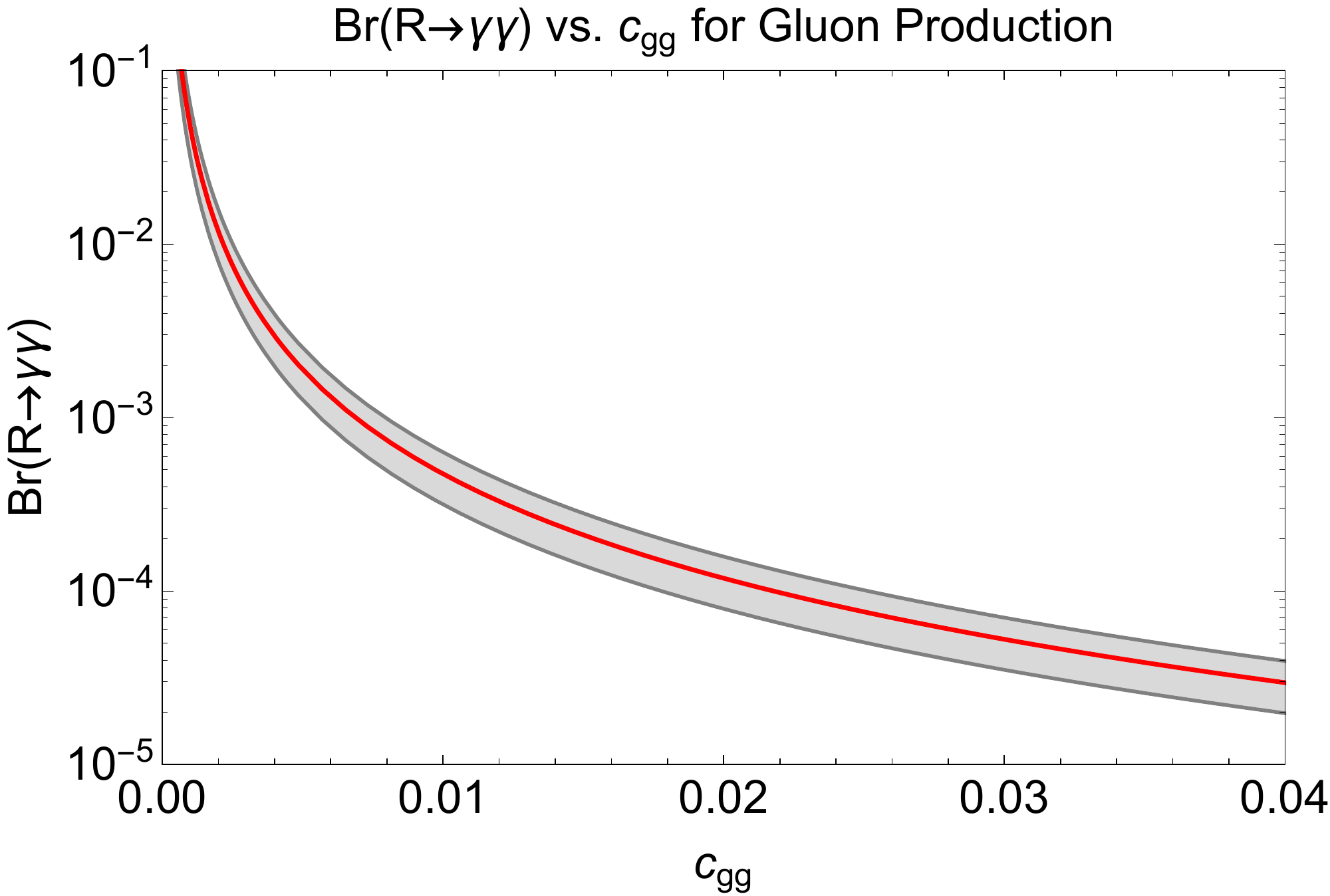}
\caption{The required relationship between the branching ratio of $R \rightarrow \gamma \gamma$ and gluon coupling $c_{gg}$ to match the observed event rate, varied between 3-6 fb, assuming gluon fusion dominates. The central value (red) corresponds to a 4.5 fb production cross section.} \label{fig:BRvsCgg}
\end{figure}

For comparison to the gluon fusion scenario, the effective operator responsible for production is
\begin{equation}
\frac{c_{gg}}{v} R G^2\ .
\label{eq:operator2}
\end{equation}
The cross section at $\sqrt{s} = 13$ TeV in the narrow width approximation is given by
\begin{equation}
\sigma_{13~\text{TeV}} = 2.8\cdot10^4~\text{pb}~k~c_{gg}^2 \text{Br}(R \rightarrow \gamma\gamma)
\end{equation} 
where the $k-$factor can be approximated by matching to known NNLO results for a heavy Higgs-like scalar~\cite{Falkowski:2015swt} and is taken to be $k = 3.4$. The ratio $\sigma_{13~\text{TeV}}/\sigma_{8~\text{TeV}}$, determined from {\tt MadGraph}, for $ggF$ is 4.5. Matching the diphoton cross section to the observed excess event rate determines the relationship between $\text{Br}^2(R \rightarrow \gamma\gamma)$ and $c_{gg}$ as shown in Fig.~\ref{fig:BRvsCgg}.

\section*{Observable Effects of the Production Proccess}

The production mechanism of a 750 GeV resonance can be probed by measuring the hadronic observables in the diphoton signal events. The most prominent features in the $\gamma \gamma$F events compared to $gg$F are the suppression of central jets with $|\eta| < 4$ and the appearance of central pseudorapidity gaps. These effects are absent for a resonance produced via $gg$F because the two incoming protons are color-connected, and the color flux tube breaking fills up the central region with soft hadrons from the fragmentation. However, for a $t$-channel exchange of a color-singlet, as in photoproduction, the remnants from each initial proton remain color singlets and only form color connections with partons originating within the same proton, typically resulting in very forward jets (or intact protons) with a pseudorapidity gap in the central region. The two forward jets expected in $W/Z$ fusion, from the parton recoil after emitting the heavy gauge boson, are typically not detected in $\gamma\gamma$F production, although two forward jets would be a signature of a resonance produced via $W/Z$ fusion.

These observables have been well-studied for VBF Higgs production~\cite{Dokshitzer:1991he, Lungov:1995iq, DeRoeck:2002hk}. For the Higgs, $gg$F dominates the cross section, making the hadronic observables of the VBF events difficult to measure. For a 750 GeV photoproduced diphoton resonance, however, the event color flow effects of the VBF production should be observable since the small background rate in the diphoton channel does not bury the signal.
	
One experimental challenge is then to remove pileup which contaminate the events with central hadronic activity. On average, there are between 10-20 inelastic collisions per bunch crossing at 2015 Run II luminosities. ATLAS and CMS have powerful tracking capabilities making possible the matching of jets to the hard interaction and identification of pileup jets~\cite{CMS:2014ata,ATLAS:2014cva}. To remove the jets resulting from pileup, only the jets matching the primary vertex of the hard interaction should be considered. CMS has shown the efficiency to match jets with $p_T > 20$ GeV to the hard interaction to be $\gtrsim 90\%$ in~\cite{CMS:2014ata} and to be flat over the range $- 2.5 < \eta < 2.5 $, corresponding to the rapidity acceptance of the tracker. Pileup tracks must also be subtracted from the event in order to observe central rapidity gaps. Only tracks matched to the diphoton primary vertex, perhaps by a longitudinal impact parameter cut, should be considered in determining whether an event contains a pseudorapidity gap. This pileup subtraction procedure may be non-standard, and the presence of central pileup tracks not matched to the primary vertex could bury this signal if special care is not taken.

In this section, we present the results of the event generation and fast detector simulation for the two production scenarios of a 750 GeV scalar or tensor diphoton resonance and the dominant irreducible diphoton background at $\sqrt{s} = 13$ TeV. The effective operators in Eq.~\eqref{eq:operator} and Eq.~\eqref{eq:operator2} have been implemented in model UFO files created with {\tt FeynRules}~\cite{Alloul:2013bka}. We also simulate the production of a spin-2 resonance $S_{\mu \nu}$ coupled to the stress energy tensor $T_{\mu \nu}$:
\begin{equation}
	\frac{c_{\gamma\gamma}}{\Lambda} S^{\mu \nu} T^{\gamma\gamma}_{\mu \nu} + \frac{c_{gg}}{\Lambda} S^{\mu \nu} T^{gg}_{\mu \nu},
\end{equation}
where $T^{\gamma\gamma}_{\mu\nu}=F_{\mu\alpha}F_{\nu\beta}g^{\alpha \beta} - \frac{1}{4} g_{\mu\nu}F_{\alpha\beta}F^{\alpha\beta}$ and the same definition for $T^{gg}_{\mu\nu}$ but with $F$ replaced by $G$.  Simulation of the hard process is performed at LO with {\tt MadGraph 5}~\cite{Alwall:2014hca}, followed by parton showering with {\tt Pythia 8}~\cite{Sjostrand:2007gs}. We use {\tt Delphes 3}~\cite{deFavereau:2013fsa} for the fast detector simulation with the default ATLAS detector geometry and efficiencies.

The irreducible, prompt $\gamma \gamma$ background is simulated at next-to-leading order (NLO) using {\tt MadGraph@NLO} followed by {\tt Herwig 6}~\cite{Corcella:2000bw}. At tree level, the only contribution is $q \bar{q}$ annihilation. Including NLO QCD corrections, this background is the dominant source accounting for more than $90\%$ ($80\%$) of the background near the excess invariant mass region in the EBEB (EBEE) selection category for the CMS diphoton search~\cite{CMS:2015dxe}. The reducible background consists of $\gamma j$ and $jj$ events in which jet fragments are misidentified as photons. We expect similar background composition for the ATLAS diphoton search. 

We define an accepted diphoton event for both signal and background as an event containing:
\begin{itemize}
\item{two reconstructed photons,}
\item{leading photon $p_T > 35$ GeV and subleading $p_T > 20$ GeV,}
\item{ 725 GeV $< m_{\gamma \gamma} <$ 775 GeV, and}	
\item{each photon satisfies isolation requirements within a $\Delta R =0.5$ cone.}
\end{itemize}
This invariant mass window is conservative, and the signal to background ratio could be improved with a tighter cut on invariant mass around the resonance.
In the accepted diphoton events, tracker and calorimeter information is used to cluster jets with {\tt FastJet}~\cite{Cacciari:2011ma} according to the anti-$k_T$ algorithm (size parameter $\Delta R = 0.6$ and minimum jet $p_T = 20$ GeV).

\begin{figure}[t!]
\center
\includegraphics[width=.49\textwidth]{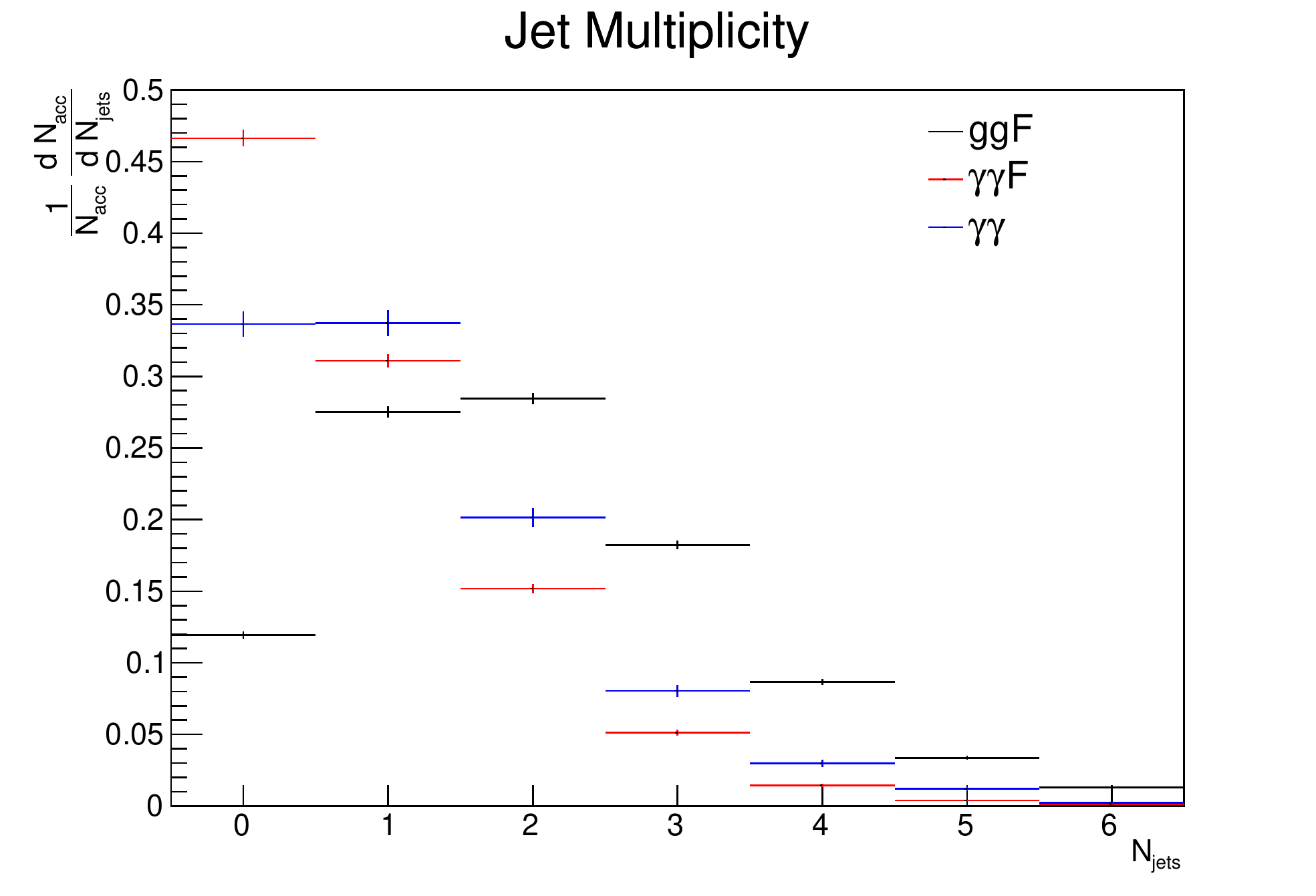}
\hfill
\includegraphics[width=.49\textwidth]{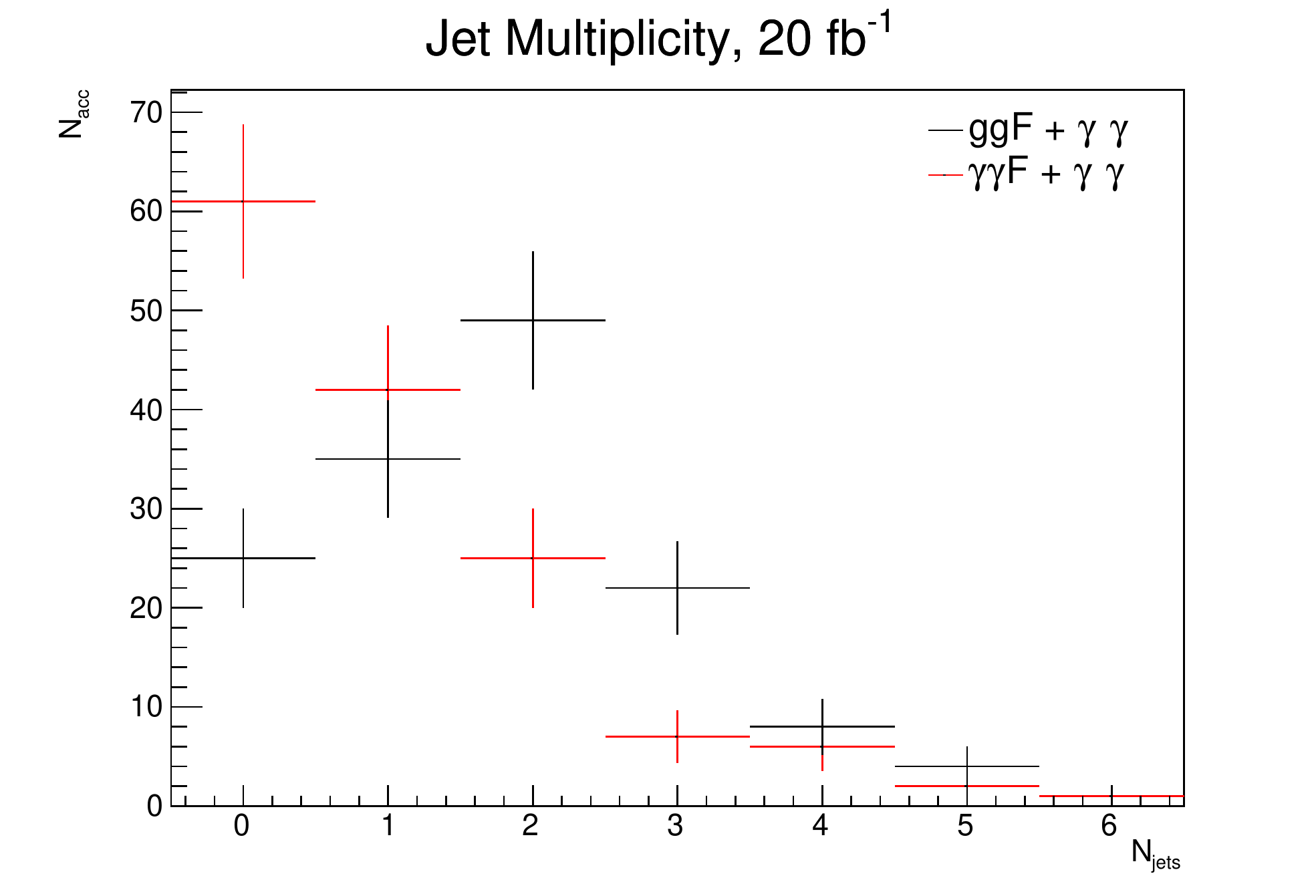}
\caption{Left: jet multiplicity per accepted diphoton event for $gg$F (black), $\gamma \gamma$F (red), and irreducible $\gamma \gamma$ background (blue). Right: sample of accepted $gg$F signal (black) and $\gamma \gamma$F signal (red) both combined with 50\% $\gamma \gamma$ background contamination with statistics corresponding to 20 fb$^{-1}$. Uncertainties are statistical only.} \label{fig:njets}
\end{figure}

\begin{figure}[t!]
\center
\includegraphics[width=.49\textwidth]{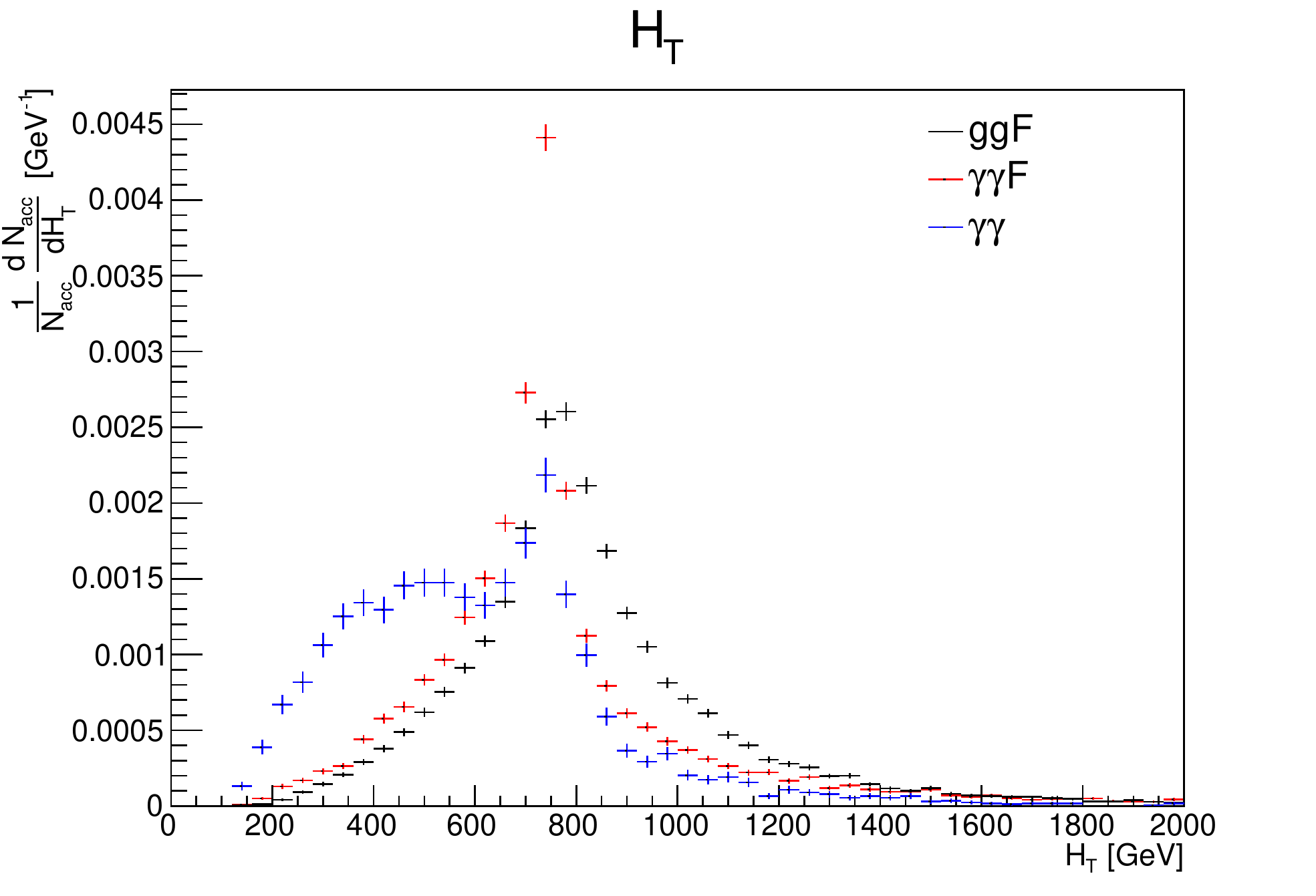}
\hfill
\includegraphics[width=.49\textwidth]{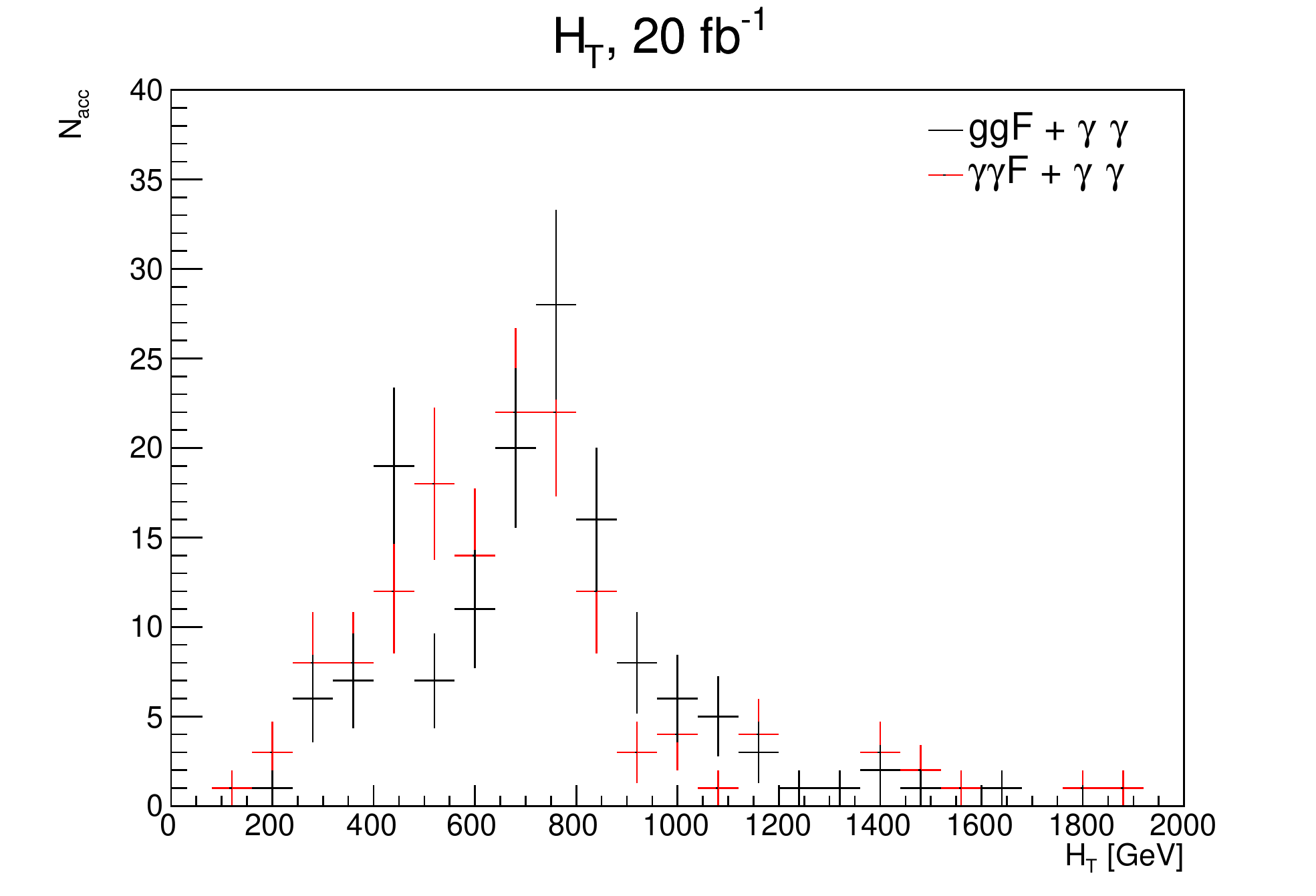}
\caption{Left: $H_T$ distribution per accepted diphoton event for $gg$F (black), $\gamma \gamma$F (red), and irreducible $\gamma \gamma$ background (blue). Right: sample of accepted $gg$F signal (black) and $\gamma \gamma$F signal (red) both combined with 50\% $\gamma \gamma$ background contamination with statistics corresponding to 20 fb$^{-1}$. Uncertainties are statistical only.} \label{fig:ht}
\end{figure}

\begin{figure}[]
\center
\includegraphics[width=.49\textwidth]{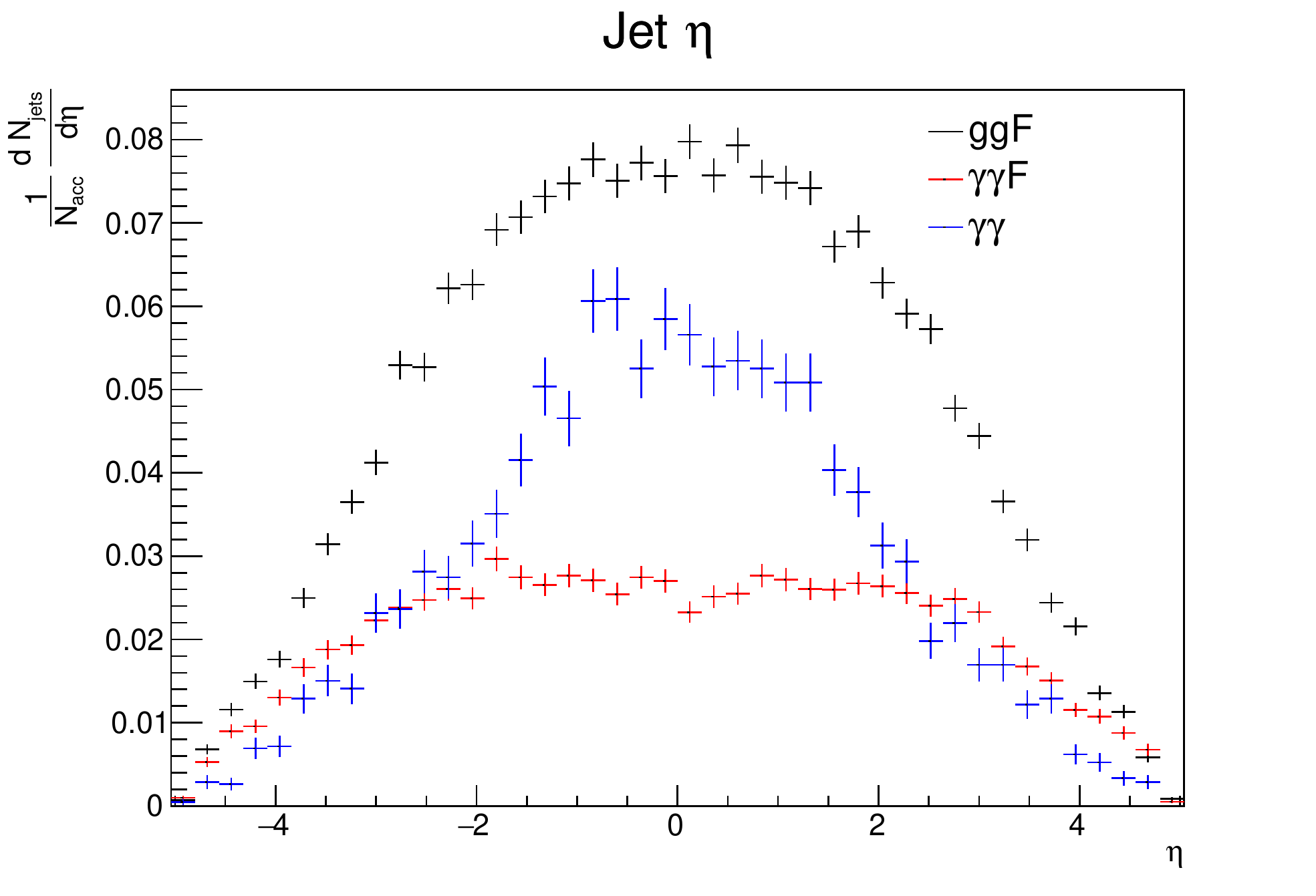}
\hfill
\includegraphics[width=.49\textwidth]{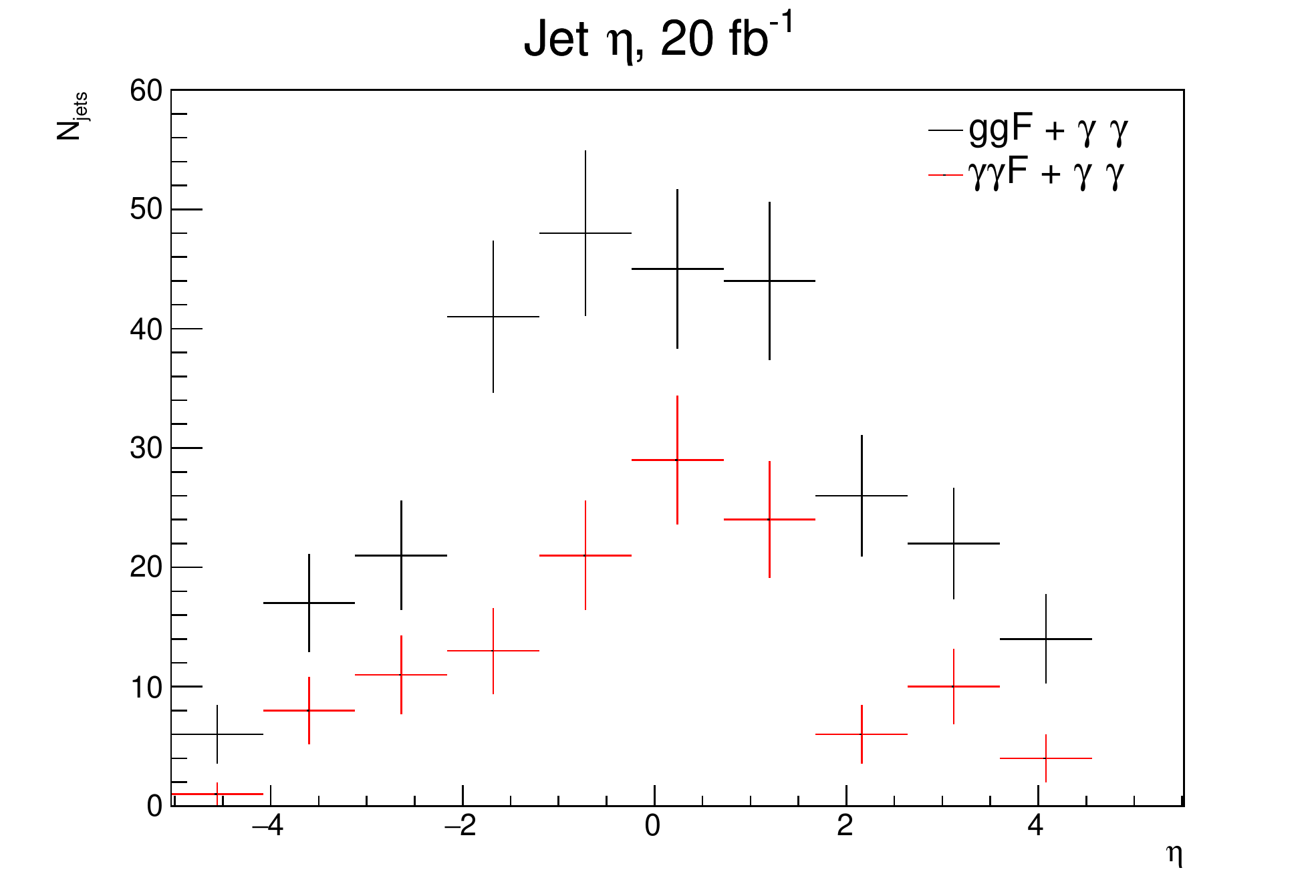}
\caption{Left: jet pseudorapidity distribution per accepted diphoton event for $gg$F (black), $\gamma \gamma$F (red), and irreducible $\gamma \gamma$ background (blue). Right: sample of accepted $gg$F signal (black) and $\gamma \gamma$F signal (red) both combined with 50\% $\gamma \gamma$ background contamination with statistics corresponding to 20 fb$^{-1}$. Uncertainties are statistical only.
} \label{fig:jeteta}
\end{figure}

\begin{figure}[t!]
\center
\includegraphics[width=.49\textwidth]{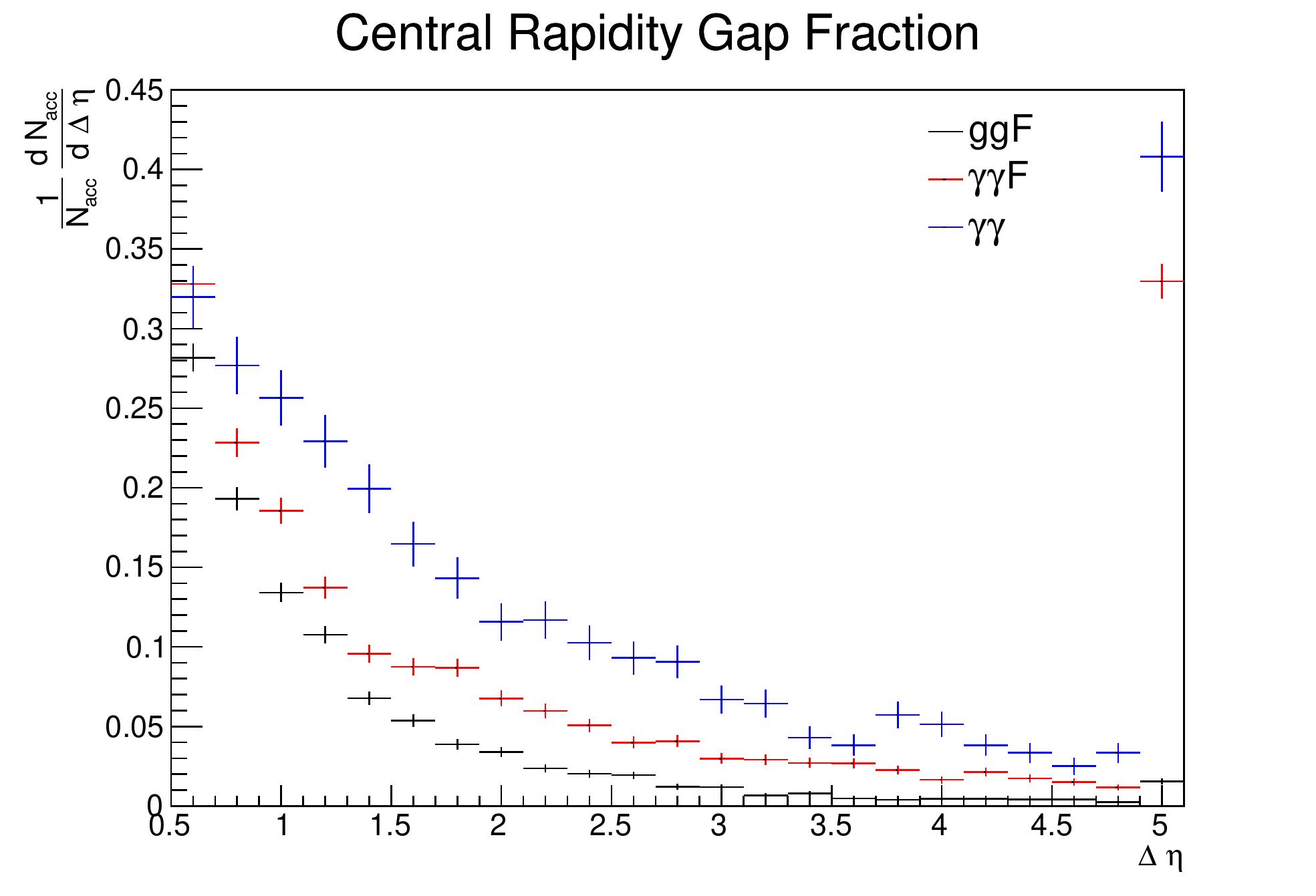}
\hfill
\includegraphics[width=.49\textwidth]{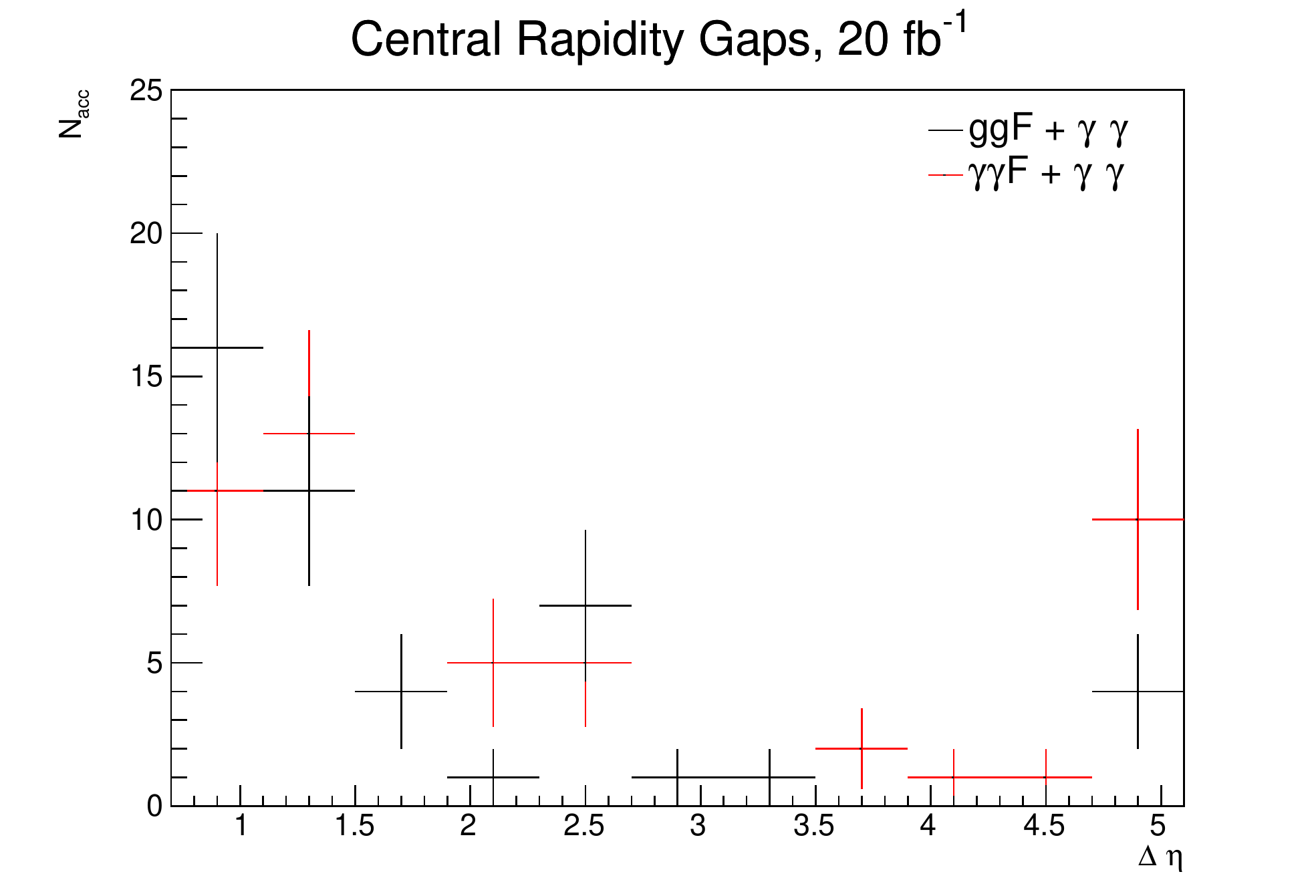}
\caption{Left: fraction of accepted diphoton events containing a central pseudorapidity gap (see text for definition) of size $\Delta \eta$ for $gg$F (black), $\gamma \gamma$F (red), and irreducible $\gamma \gamma$ background (blue). Right: sample of accepted $gg$F signal (black) and $\gamma \gamma$F signal (red) both combined with 50\% $\gamma \gamma$ background contamination with statistics corresponding to 20 fb$^{-1}$. Uncertainties are statistical only.} \label{fig:gaps}
\end{figure}

\begin{figure}[t!]
\center
\includegraphics[width=.49\textwidth]{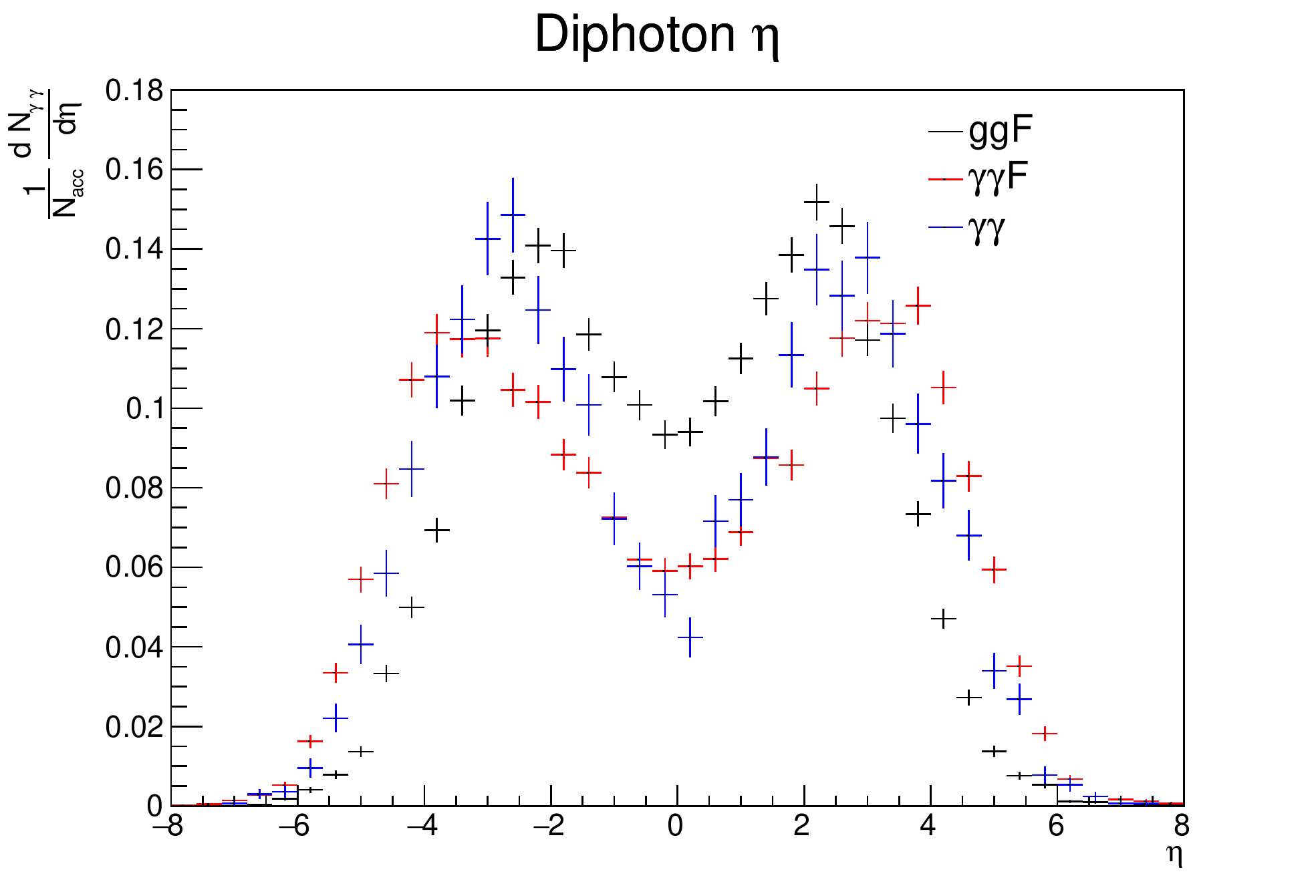}
\hfill
\includegraphics[width=.49\textwidth]{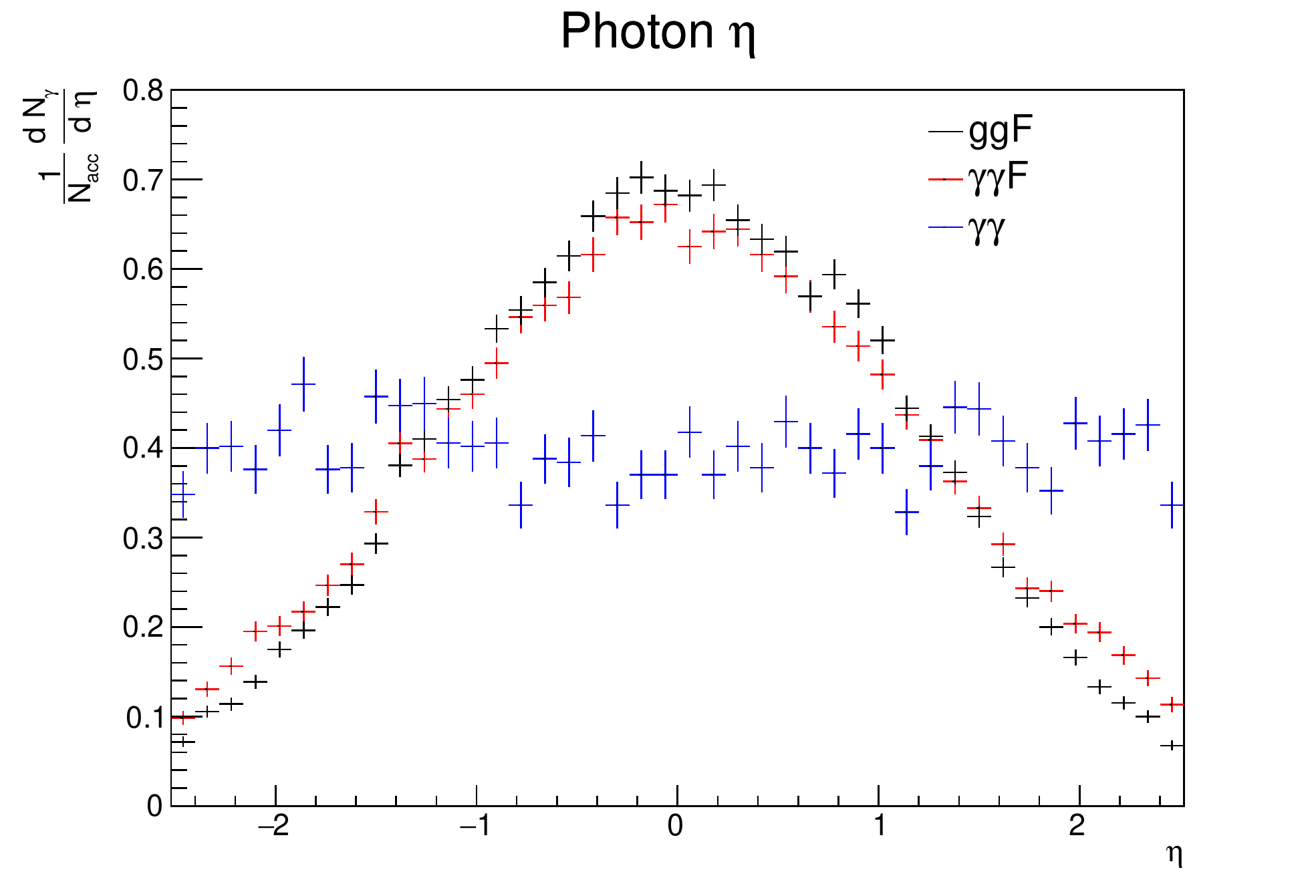}
\caption{Left: pseudorapidity of accepted diphotons for $gg$F (black), $\gamma \gamma$F (red), and irreducible $\gamma \gamma$ background (blue). Right: pseudorapidity of photons in accepted diphoton events, normalized by the number of accepted diphoton events, $N_{acc}$, for $gg$F (black), $\gamma \gamma$F (red), and irreducible $\gamma \gamma$ background (blue). Uncertainties are statistical only.} \label{fig:eta}
\end{figure}

The conclusions drawn from hadronic observables concerning the resonance production mechanism for the scalar signal also apply to the tensor case, so we only present hadronic observables of the scalar signal. In Figs.~\ref{fig:njets} and ~\ref{fig:ht} (left), we present the jet multiplicity and scalar sum of transverse energy, $H_T$, of accepted diphoton events for $gg$F and $\gamma\gamma$F scalar prodution and irreducible background $(\gamma\gamma)$. These observables demonstrate the additional hadronic activity, also visible in the charged particle multiplicity, in $gg$F events compared to $\gamma \gamma$F or background events. The $gg$F events prefer 2.0 jets per event, while the $\gamma \gamma$F and irreducible $\gamma\gamma$ events have an average of 0.9 and 1.3 jets per event. However, a $gg$F scenario would not be distinguishable with the existing data due to background contamination. We expect a 5-$\sigma$ statistically significant excess in the total number of jets for the case of a $ggF$ produced resonance at 10 fb$^{-1}$ of data. The peak in the number of events for both signals and background at $H_T \sim 750$ GeV is due to the energy of the diphoton. For a $gg$F produced resonance, the $H_T$ distribution of signal events would contain excess events in the $H_T > 800$ GeV region, which can be attributed to the extra jets in $gg$F events. Also in Figs.~\ref{fig:njets} and ~\ref{fig:ht} (right), we present the same distributions for a for a $50\%$-signal, $50\%$-$\gamma \gamma$ sample of events with statistics corresponding to 20 fb$^{-1}$. The difference in jet multiplicity for $gg$F vs. $\gamma\gamma$F production is clearly discernible at 20 fb$^{-1}$.

In additional to the difference in overall amount of hadronic activity in the two production scenarios, there is also different angular dependence due to the event color flow and kinematics of the lab frame. In Fig.~\ref{fig:jeteta} (left), we show the suppression of central jets for $|\eta_j| \lesssim 4.5$ in $\gamma\gamma$F events. This effect is somewhat washed out due to the peak in background $\gamma \gamma$ events at central jet rapidities.

The most striking feature comparing $\gamma \gamma$F to $gg$F events is the appearance of central pseudorapidity gaps for the former. We define a central rapidity gap as the maximum symmetric region of pseudorapidity around $\eta = 0$ for which there are no tracks with $p_T > p_{T,min}$. We choose to only consider tracks in the determination of this observable in order to guarantee the pileup contamination can be removed by tracker information. The ATLAS and CMS trackers have acceptance out to $|\eta| < 2.5$, giving a maximum observable (track) gap size of $\Delta \eta$ = 5. The fraction of signal events containing such a rapidity gap for $p_{T,min} = 1$ GeV is shown in Fig.~\ref{fig:gaps} (left). The choice of $p_{T,min}$ was made in order to optimize the difference in number of gaps between $gg$F and $\gamma \gamma$F signals and to ensure the tracks are high quality in order to be matched to the diphoton vertex, but we note that $p_{T,min} \in [0.8,2]$ GeV is useful for discriminating between $gg$F and $\gamma \gamma$F production. The number of events in the $\Delta \eta = 5$ bin is large from overflow of events with larger gap sizes (which cannot be measured due to the acceptance of the tracker). Large rapidity gaps are exponentially suppressed for $ggF$ events, so a $gg$F signal component would decay exponentially with $\Delta \eta$ compared to (approximately) linearly fall of the background with $\Delta \eta$. The fraction of accepted diphoton events containing a $\Delta \eta \geq 3$ gap are 1.6\%, 11.7\%, and 19.1\% for $gg$F, $\gamma \gamma$F, and $\gamma \gamma$ background events, respectively. This suggests an enhancement to the rapidity gap rate for the case of $\gamma \gamma$F production and a suppression compared to background for $gg$F. This difference in the predicted number of rapidity gaps in the two scenarios is not statistically significant with the existing data. In Fig.~\ref{fig:gaps} (right), we show a sample of central rapidity gaps corresponding to 20 fb$^{-1}$ of data. One caveat with measuring rapidity gaps is that it relies on modeling of nonperturbative effects and cannot be precisely predicted perturbatively.


\begin{figure}[t!]
\center
\includegraphics[width=.49\textwidth]{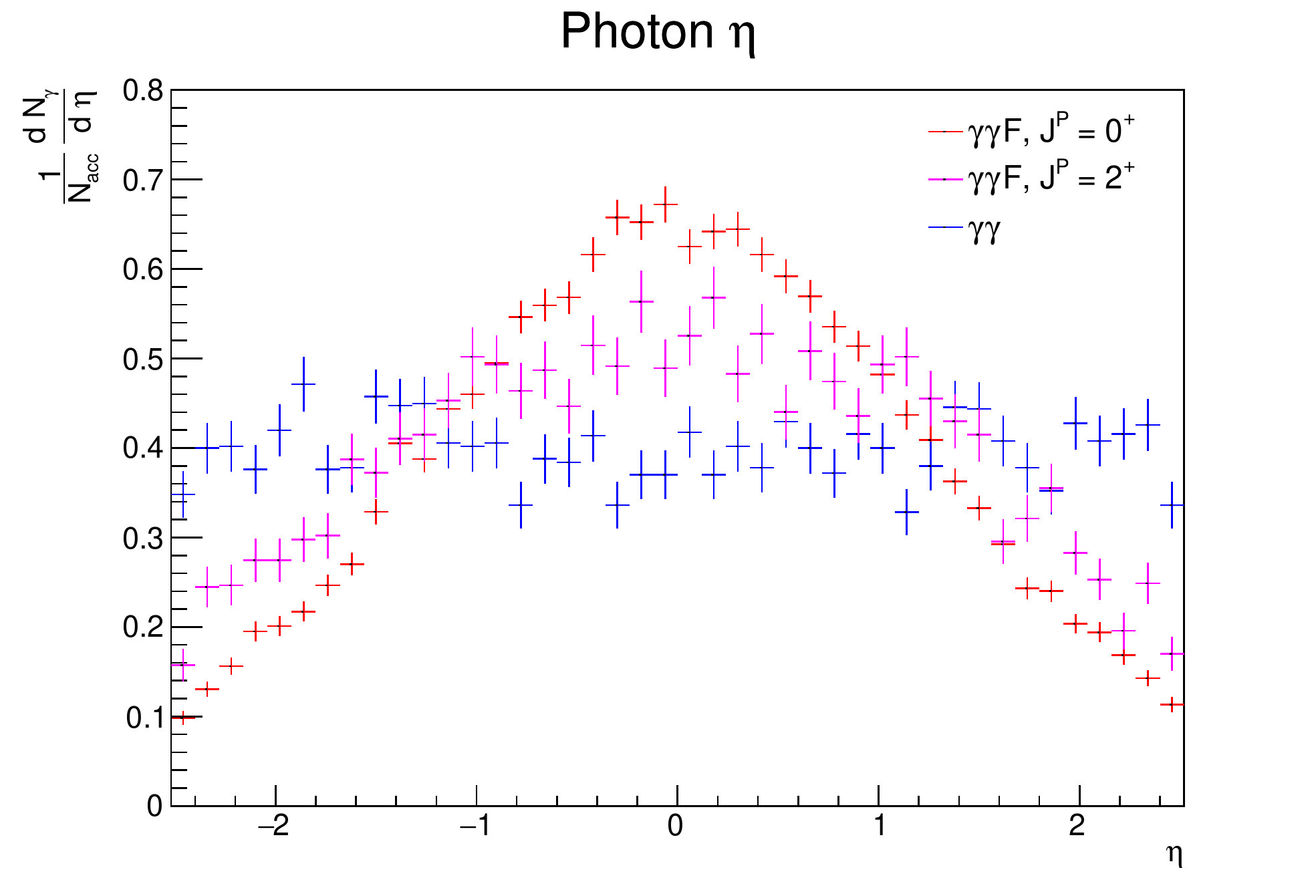}
\hfill
\includegraphics[width=.49\textwidth]{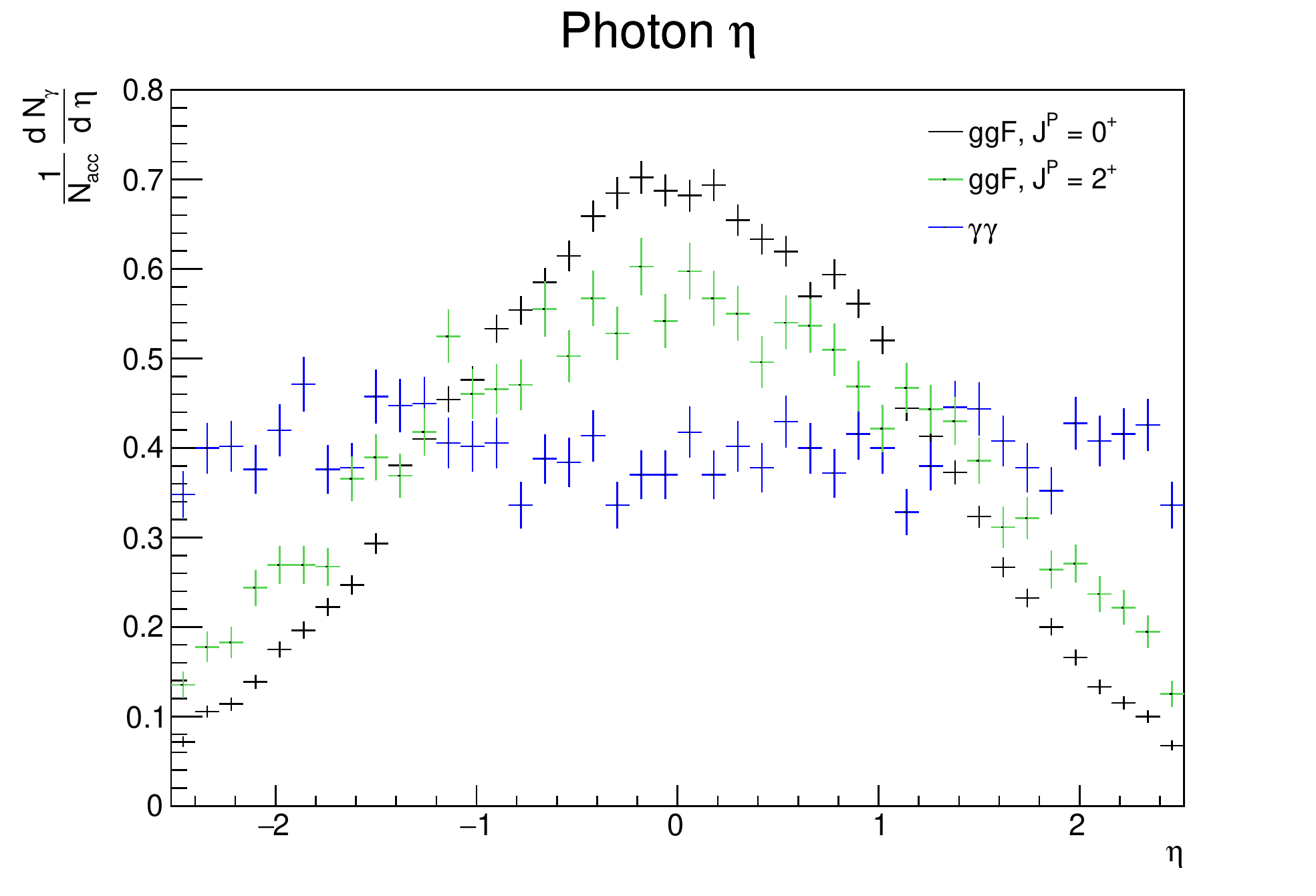}
\caption{Left: pseudorapidity of photons in accepted diphoton events, normalized by the number of accepted diphoton events, $N_{acc}$, for $\gamma\gamma$F production of a spin-0 (red) or spin-2 (purple) resonance and irreducible $\gamma \gamma$ background (blue). Right: pseudorapidity of photons in accepted diphoton events for $gg$F production of a spin-0 (black) or spin-2 (green) resonance and irreducible $\gamma \gamma$ background (blue). Uncertainties are statistical only.} \label{fig:costheta}
\end{figure}

The two production modes considered also have different parton distribution functions and kinematics affecting the hard process in the events. We show these effects in the diphoton and photon pseudorapidity distributions in Fig.~\ref{fig:eta}. The diphoton (resonance) tends to be more forward produced in $\gamma \gamma$F events, peaking at $|\eta_{\gamma\gamma}| \simeq 3.5$. In $gg$F signal events, $\eta_{\gamma\gamma}$ peaks at $|\eta_{\gamma\gamma}| \simeq 2.5$, while the dip in diphoton production at $\eta_{\gamma \gamma} = 0$ for $gg$F events is not as pronounced. The rapidity distribution of the individual photons in Fig.~\ref{fig:eta} (right) is flat for the background but peaked at the center of the detector for both signals suggesting that a scalar diphoton resonance prefers to decay to photons in the detector barrel. This can be understood as follows. In events where a heavy scalar resonance is produced, the decay photons are isotropic in the parton center of mass frame. In the lab frame, the boost factor of the resonance is typically small resulting in an almost isotropic $\cos{\theta_{\gamma}}$ distribution (with small peaks towards $\cos{\theta_{\gamma}}=\pm 1$ from the small boost factor of the resonance along the beam axis) for the decay photons. The shape of the $\eta_{\gamma}$ distribution is dominated by the Jacobian factor from changing variables from $\cos{\theta_{\gamma}}$ to $\eta_{\gamma}$, which has the same qualitative shape as the photon pseudorapidity distributions for the resonance signals. The $\gamma\gamma$ background events, on the other hand, are highly boosted with a $\cos{\theta_{\gamma}}$ distribution that is strongly peaked at $\cos{\theta_{\gamma}} = \pm 1$. When combined with the Jacobian factor, the $\eta_{\gamma}$ distribution of the $\gamma \gamma$ events turns out to be flat. This is not the case for a spin-2 resonance produced via $gg$F or $\gamma\gamma$F, which does not decay isotropically to photons in the parton center of mass frame~\cite{Boer:2013fca, ATLAS:2013xla, Han:2015cty} resulting in more forward photons than a scalar signal. In Fig.~\ref{fig:costheta}, we present the $\eta_{\gamma}$ distributions for $gg$F and $\gamma\gamma$F production of a spin-0 and spin-2 resonance, confirming that a tensor resonance gives more forward photons than the scalar signal.


\begin{table}[t!]\centering
\begin{tabular}{lccccc}
\toprule 
\multirow{2}{*}{CMS Diphoton Selection} & \multicolumn{2}{c}{\boldmath$gg$\bfseries F} & \multicolumn{2}{c}{\boldmath $\gamma\gamma$\bfseries F} & \multirow{2}{*}{\boldmath $\gamma\gamma$}\\
\cmidrule(l){2-3} \cmidrule(l){4-5}
 & $J^P = 0^+$ & $J^P = 2^+$ & $J^P = 0^+$ & $J^P = 2^+$ \\
\midrule
EBEB & $72.8\%$ & $58.2\%$ & $70.0\%$ & $55.2\%$ & $60.0\%$ \\
EBEE & $27.2\%$ & $41.8\%$ & $30.0\%$ & $44.8\%$ & $40.0\%$ \\
\bottomrule
\end{tabular}
\caption{Expected fraction of diphoton events accepted by the CMS diphoton search~\cite{CMS:2015dxe} with both photons detected in the barrel (EBEB category) or with one photon in the barrel and the other in the end cap (EBEE category) for $gg$F and $\gamma\gamma$F production of either a spin-0 or spin-2 resonance along with irreducible $\gamma\gamma$ background events.}
\label{tab:tab1}
\end{table}

CMS reported an unexpectedly large number of the excess events in the EBEE selection category, in which one of photons from the diphoton is detected in the end cap ($|\eta_{\gamma}| > 1.57$) and one in the barrel ($|\eta_{\gamma}| < 1.44$), rather than both photons detected in the barrel (EBEB category)~\cite{CMS:2015dxe}. Whether these events should be considered as signal events depends on the width and mass of the resonance. We apply the cuts from the CMS diphoton analysis event selection to obtain the expected number of events in the EBEE and EBEB categories for the signal and background in Table~\ref{tab:tab1}. The signal diphoton events (for both a $gg$F or $\gamma\gamma$F produced scalar resonance) are most likely to be accepted in the EBEB selection category with $\gtrsim 70\%$ of accepted signal events expected in this category. This is a direct consequence of the $\eta_{\gamma}$ distributions in Fig.~\ref{fig:eta} (right) for the $gg$F and $\gamma\gamma$F signals. For the spin-2 signal, however, more events are accepted in the EBEE category compared to the scalar signal for both $gg$F and $\gamma\gamma$F production. The irreducible background diphotons are more likely to show up in EBEE category than the photons from a 750 GeV scalar resonance. We note that the expected number of background events reported by CMS in the EBEE category is larger than the number expected in the EBEB category, which is not the case for the irreducible $\gamma\gamma$ background alone. We understand this effect to be due to be the additional presence of $jj$ background events, which we do not simulate, that account for $\sim 20\%$ of the EBEE category expected background but a significantly smaller fraction for the EBEB category.

\section*{Conclusions} 

We have simulated the two production mechanism of a hypothetical 750 GeV diphoton resonance at LO and the largest source of irreducible diphoton background at NLO. The main differences between the two production modes are the jet multiplicities and jet rapidities of the accepted diphoton events in addition to the intact protons from elastic photoproduction, all which may be observed with high statistical significance around or before 20 fb$^{-1}$ of data.  More subtle features of the signal events include an excess number of events with $H_{T} > 800$ GeV for $ggF$ production and an enhancement to the central pseudorapidity gap rate for $\gamma\gamma$F production. Moreover, we find that scalar resonance signals strongly prefer to decay to photons in the barrel rather than the endcap, hinting that, for a wide resonance interpretation, the large fraction of excess events accepted in the EBEE diphoton selection category by the Run-2 CMS diphoton search may not be explained by a spin-0 resonance. If the excess persists with the same features after the collection of more data, a spin-2 resonance may better describe the number of photons detected in the detector end cap.

\section*{Acknowledgments}

We would like thank Jordan Tucker, Nathan Mirman, and Wee Hao Ng for useful discussions. C.C.~and S.L~are supported in part by the NSF grant PHY-1316222.  J.H.~is supported in part by the DOE under grant DE-FG02-85ER40237.  J.T.~is supported in part by the DOE under grant DE-SC-000999.


%

\end{document}